# Facile and ultraclean graphene-on-glass nanopores by controlled electrochemical etching

Xiaoyan Zhang[a], Pauline M.G. van Deursen[a], Wangyang Fu[b,*], Grégory F. Schneider[a,*]

[a]*Leiden Institute of Chemistry, Leiden University, Einsteinweg 55, 2333 CC Leiden, Leiden, Netherlands*

[b]*School of Materials Science and Engineering, Tsinghua University, No.1 Tsinghua Yuan, Haidian District, 100084 Beijing, China*

*Corresponding authors.

 *E-mail addresses*: fwy2018@mail.tsinghua.edu.cn (W. Fu),

 g.f.schneider@chem.leidenuniv.nl (G. Schneider)

**ABSTRACT:**

A wide range of approaches have been explored to meet the challenges of graphene nanostructure fabrication, all requiring complex and high-end nanofabrication platform and currently suffering from surface contaminations, which not only give rise to potential electrical noise but also hinder its ideally atomic spatial resolution. Here, with the use of an electrical pulse on a low-capacitance graphene-on-glass (GOG) membrane, we fabricated ultraclean graphene nanopores on a commercially available glass substrate with exceptionally low electrical noise. *In-situ* liquid AFM studies and electrochemical measurements revealed that both graphene nanopore nucleation and growth are stemming from the electrochemical attack on carbon atoms at defect sites, ensuring an inherently ultraclean graphene nanopore. Strikingly, compared to conventional TEM drilled graphene nanopores on SiN supporting membranes, our developed GOG nanopores featured with an order-of-magnitude lowered broadband noise, which we ascribed to the electrochemical refreshing of graphene nanopore on mechanically stable glass chips with negligible parasitic capacitance (~1 pF). Further experiments on double-stranded DNA translocations not only demonstrated the greatly reduced current noise, but also confirmed the formation of single GOG nanopores. Therefore, the exceptionally low noise and ease of ultraclean GOG nanopore platform will facilitate innovative researches to understand the fundamental property and application of such atomically thin nanopore sensors.

**KEYWORDS:** graphene nanopores, ultraclean, low noise, electrochemical etching, glass substrate

Owing to the graphene atomically thin and ion-impermeable structure, nanopores[1-3] and nanogaps[4] in graphene provide the opportunity for reaching single building block resolution in biomolecule sequencing.[5] In nanopore sequencing, a strand of DNA traverses a nanopore submerged in a saline recording solution. The ionic current through the pore is monitored as the DNA strand passes, providing information on the local diameter of the strand. From the standpoint of device fabrication, the challenge lies in the reliable formation of atomic clean and stable pores or gaps in graphene at the nanometer scale.[6] Among the current methodologies, transmission electron microscopy (TEM) sculpting with a highly focused electron beam is the most controllable.[7] Nevertheless, the high costs and potential carbon contamination of the TEM technique makes it unsuitable for upscaling the production of nanopore or nanogap devices. In addition, the TEM processing step limits the choice of substrates, as TEM substrates need to be compatible with a TEM holder and electrically conductive – to prevent charge accumulation during exposure to the electron beam. Free-standing thin silicon-nitride (SiN) membranes (several tens of nanometers in thickness supported on Si substrates) with a pre-fabricated nanopore (several hundred of nanometers in diameter), are commonly used as supporting substrates for TEM-fabricated graphene nanopores.[2, 3] The presence of a Si chip, however, raises a drawback during the nanopore measurement when submerged in the saline solution: large parasitic capacitance (»10-100 pF) resulting from the relatively thin supporting SiN membrane between the semiconducting Si and the recording solution induces high current noise.[8, 9] Recently, electrical pulse fabrication was used as an alternative route to efficiently and reproducibly fabricate graphene nanopores.[10] However, so far such graphene nanopores were fabricated on silicon-based semiconducting substrates with relatively high current noise and low bandwidth. Particularly, to a large extend, the mechanism of nanopore formation on graphene with the electrical pulse method, as well as the ultraclean nature of the nanopores, are unclear and therefore seriously limiting the applications of the electrical pulse approach.[10]

In this paper, we use low-capacitance, commercially available glass chips featuring a micrometer-sized aperture at the center and use it as a support for graphene. This graphene-on-glass (GOG) platform, as we called it, is then subjected to electrically pulsed voltages yielding the formation of an ultraclean graphene nanopore. In contrast to the dielectric breakdown fabrication, we reveal a novel electrochemical mechanism using *in-situ* liquid AFM, where a graphene nanostructure is enlarged by removing carbon atoms at its freestanding edges during electrical fabrication. Further experiments on double-stranded DNA translocations not only confirmed the formation of a single GOG nanopore, but also demonstrated greatly reduced current noise which we ascribed to the electrochemical refreshing of graphene nanopore on mechanically stable glass chips with minimized parasitic capacitance.

**EXPERIMENTAL METHOD**

**CVD graphene transfer and liquid handlings.** Polycrystalline CVD graphene was obtained from Graphenea with good monolayer homogeneity (>98 %) and micrometer-sized grains. Raman spectra and imaging were performed on a WITec confocal spectrometer with 100×objective (lateral resolution of ~300 nm) and blue laser (488 nm, 1 mW), confirms the relatively high quality of graphene on copper foils (not shown). Prior to CVD graphene transfer, the Nano-Patch-Clamp chips (Nanion) were subjected to sonication in DI water, ethanol, and followed by blow drying with nitrogen. The glass chips were then treated in an oxygen plasma for 2 mins. Capacitively coupled plasma system with the radio-frequency (RF) of 40 kHz and 200 W power from Diener electronic (Femto) was employed at room temperature with a base pressure <0.02 mbar and 50 W/2.4 mbar for oxygen plasma. PMMA supported CVD graphene layer floating on pure DI water (which was achieved by spin coating PMMA, copper etching, and DI water rinsing) was then transferred onto such prepared glass chips. The PMMA layer was dissolved in acetone in the last step of transfer, leaving freestanding CVD graphene on the glass chips with micrometer-sized conical shaped aperture at the centre (Fig. 1). We adopted a protocol to use 50 % ethanol (in DI water), DI water, and 1 M KCl/10 mM Tris solution (~11 S/m at room temperature) at pH 8.1. Such protocol allows completely wetting and electrical contact via Ag/AgCl electrodes to both sides of the graphene membrane, as any air bubbles will be repelled because of the low surface tension of the 50 % ethanol solution. Normally intact CVD graphene membranes display a very small leakage conductance (<1 nS), which is negligible compared to pore conductance (>10 nS).

**AFM microscopy.** AFM images were performed on a JPK NanoWizard Ultra Speed machine. A silicon probe (AC240TS, Asylum Research) with ~70 kHz nominal resonance frequency in air and ~30 kHz resonance frequency in liquid, was used. The images were scanned in an intermittent contact mode, which allows to work with softer and easy to damage material, in liquid and at room temperature with $512 \times 512$ pixels. Such collected images were then processed using a JPK SPM Data Processing software. The in-situ liquid AFM images in the inside were performed both before and after the application of electrical pulses without drying the sample. Detailed pictures and illustrations of the *in-situ* liquid AFM setup can be found in Fig. S1.

**Electrical measurements and pulse generator.** All the nanopore measurements were performed with an ultralow-noise patch clamp amplifier (Axopatch™ 200B, Molecular Devices) with resistive-feedback circuitry in whole-cell mode. Ag/AgCl electrodes are used to detect ionic currents and to apply electric fields. An Agilent pulse generator (81101A) was used to apply 50 ms long voltage pulses. To protect the amplifier from large input voltages, a mechanical switcher was used to disconnect the amplifier before each pulse. The Axopatch™ 200B amplifier was reconnected 2 s after each pulse such that the GOG system gets stabilized. We empirically adopted a 7 V/5 V protocol for electrochemical etching as previous electrical pulse fabrication method also reported the lowest voltage to reliably nucleate pores (~7 V) and the lowest voltage to reliably enlarge nucleated pores (~5 V): 1) 7 V pulses with 50 ms period were applied repeatedly for both nucleation and formation, and 2) 5 V pulses with 50 ms were used for enlarging and stabilizing the GOG nanopores.

**DNA translocation and noise measurement.** λ-DNA (48,502 base pairs, Lot: 1651504, from New England Biolabs Tech. N3011S) solution was prepared into a concentration of around 15 µg mL$^{-1}$ in 1 M KCl with 10 mM tris and 1 mM EDTA, pH 8.1. DNA translocation events were recorded under ~100-200 mV bias voltage using a band pass filter of 100 kHz and a sampling rate of 500 kHz. The current traces were filtered using an 8-pole Bessel filter at 10 kHz in Clampfit. A low noise data acquisition system (Axon™ Digidata 1550A, Molecular Devices), was adopted to further eliminate local noises and to record small biological signals. Power spectral densities were calculated by taking the Fourier transform of the autocorrelation function and dividing it by the sampling frequency and the sample length. For normalization, the power spectral densities were divided by the mean current of the corresponding traces. All of the analyses and fitting were done with a combined in one single, comprehensive new Matlab GUI-based package named Transalyzer.

**RESULTS AND DISCUSSION**

As the starting substrate, commercially available glass chips (insulating borosilicate Nano-Patch-Clamp chips purchased from Nanion) were used, containing a micrometer-sized conical aperture at the center (Fig. 1a). A monolayer of chemical vapor deposition (CVD) graphene sheet was transferred onto the conical glass pore using the standard poly(methyl methacrylate) (PMMA) transfer method[11] to obtain the graphene-on-glass (GOG) platform. This platform was submerged in a saline solution (1 M KCl with 10 mM Tris and 1 mM EDTA, pH 8.1, if not specified declare) and connected on both side of the pore to an amplifier through Ag/AgCl electrodes (Fig. 1a, lower panel). In this setup, we recorded the pore resistance, capacitance and current noise spectra. Before the transfer of graphene, the 100 µm thick insulating glass substrate contributes 1.8 pF ($C_{glass}$=1.8 pF, Fig. S2) to the total input capacitance with an aperture resistance of 2.1 MΩ (Fig. 1d, gray line). Such small parasitic capacitance benefits low noise/large bandwidth sensing applications as discussed later. The as-prepared GOG ensemble exhibited a stable transmembrane resistance of 2.7 GΩ (black rectangular and green circles, Fig. 1d), indicating a complete graphene coverage of the glass pore. AFM images (Fig. 1b and c) show the morphology of the micrometer-sized glass pore with a clean, smooth-rimmed opening, before and after graphene transfer. In the 90 GOG substrates fabricated and tested, about 30 devices revealed finite transmembrane resistance in the order of 100 MΩ after transfer, attributed to inherent defects in CVD-grown graphene or electrostatic discharge effects[12] (which we will discuss later). Frequently, significant capillary force leads to damaged graphene membranes with resistance <10 MΩ. Excluding samples with damaged graphene, the yield of GOG samples for subsequent processing was 31 devices (34 %) exhibiting transmembrane resistance of >1 GΩ.

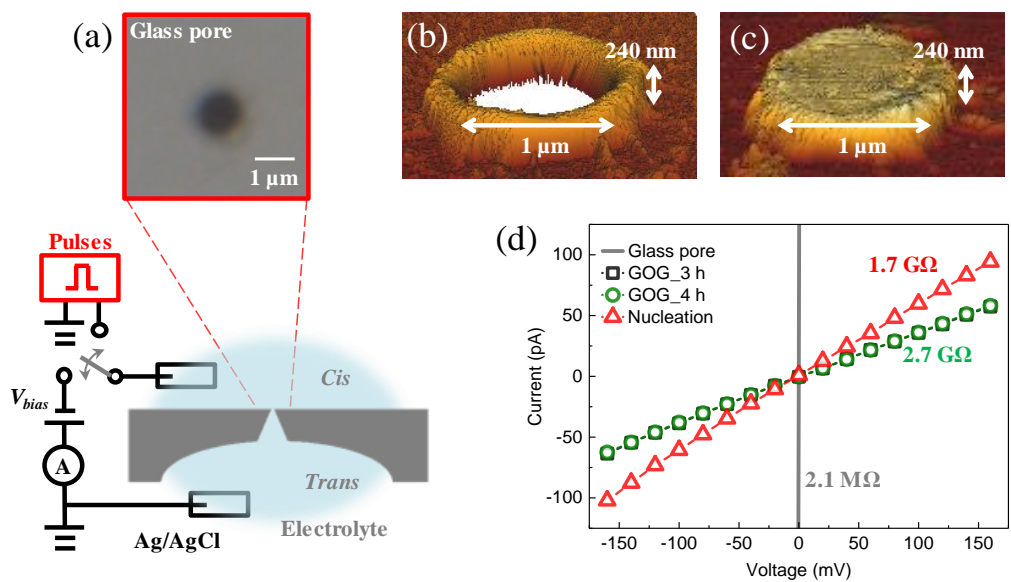

**Figure 1. Preparation and characterization of the graphene-on-glass (GOG) substrate**. (a) Upper panel: an optical image shows a well-defined aperture at the center of a glass chip; lower panel: experimental scheme consisting of "*cis*" and "*trans*" fluidic chambers with a glass pore between the two reservoirs. The Ag/AgCl reference electrode resides in the *cis* fluidic chamber was used to introduce transmembrane voltage pulses for GOG nanopore fabrication. The AFM image in (b) confirmed a clean, smooth-rimmed glass pore over a circular aperture 1 µm in diameter. (c) AFM images of a glass pore completely covered by a free-standing CVD graphene, forming a GOG substrate. (d) The IV curves of a bare glass pore (gray line) and a GOG substrate with a stable resistance of 2.7 GΩ (black rectangular and green circles) in 1 M KCl, 10 mM tris and 1 mM EDTA, pH 8.1. The corresponding increase in current (reduction in resistance, 1.7 GΩ, red triangular) suggests nanopore nucleation after applying short-time (50 ms), high-voltage (7 V) pulses.

*In situ*, a voltage was applied to open up a graphene nanopore. We applied short high voltage pulses (50 ms, 7 V) across the graphene membrane. Such a millisecond pulse is easier to achieve than the previously reported 250 ns pulses applied for graphene nanopore fabrication.[10] In Figure 1d, repeated application of high voltage pulses (50 ms, 7 V, 10 repetitions with 2 s separation) leads to a decrease in the resistance to 1.7 GΩ (red triangles). In the following, we always use 50 ms pulses with 2 s separation time, if not otherwise specified.

We formed a GOG nanopore with a resistance of 88 MΩ when continuously repeating the 50 ms pulses for another 50 times after the first opening of the pore (Fig. 3a, blue dot). To evaluate the controllability of the electrical fabrication, we averaged the relative resistance $R/R_0$ of three different GOG nanopores, which exhibit similar resistances of around 90 MΩ after experiencing 60 times 7 V/50 ms pulses. As shown in the inset of Figure 3a, after applying 7 V/50 ms pulses for 10, 20, and 60 times, the standard deviations of $R/R_0$: ($\frac{\text{error bar}}{R/R_0}$) are 22 %, 27 %, and 60 %, respectively, suggesting a reasonable reproducibility of our electrical fabrication method. In the final step, we keep enlarging and stabilizing the nanopore by using modest voltage pulses until a resistance of 54 MΩ is reached (Fig. 3a, gray dot). This resistance value of 54 MΩ corresponds to a graphene nanopore with an equivalent 2.3 nm pore diameter assuming a regular round shape.[1, 2] The fact that this resistance value is kept unchanged over a time period of at least one hour (Fig. S3), suggests that the GOG nanopore is stable.

In the case of dielectric breakdown fabrication in SiN membranes,[13] the applied electric fields (up to ~1 V nm$^{-1}$) are close to the dielectric breakdown strength of low-stress SiN films, and are able to induce accumulation of charge traps forming a highly localized conductive path. This results in the physical damage of the conductive path because of substantial power dissipation and heat generated, yielding a nanopore in the bulky SiN membrane.[13] Unlike 10-30 nm thick SiN membranes, graphene is a conducting membrane with atomic thickness. Figure 2a and Figure S4 represent the circuit diagram of the GOG and graphene on SiN systems, respectively. Instead of crossing the membrane, the potential drop $V_{appl}$ occurs mainly at the graphene/liquid interface in the glass pore where the charging capacitance is smallest: $C_{lower}$=16 fF << $C_{glass}$=1-2 pF, $C_Q$=300 pF, $C_{upper}$=60 pF, where $C_Q$ and $C_{upper}$ represent the quantum capacitance and the interface capacitance at the upper side of the graphene membrane, respectively. To date, little is known on the mechanism that is responsible for graphene nanopore nucleation and growth, and it is one of the main goals of this current study to shed light on these mechanisms.

In general, the surface of glass carries negative charges, as the main composition of glass is SiO$_2$ with –OH groups on the surface with a density of up to $5\times10^{18}$ charge per m$^2$ and an isoelectric point (IEP) around 3. The negative charge density $Q_{int}$ on the glass surface could cause critical variations $V_{vari}=Q_{int}/C_{upper}$ in the potential of the graphene membrane ($V_{ch}$), in addition to the applied potential $V_{appl}$: $V_{ch}=V_{appl}+V_{vari}$. As a rough estimation, we calculated the negatively charged surface group density $Q_{int}$ to be $5\times10^{17}$ m$^{-2}$ in a solution of 10 mM Tris (pH=8.1).[14] Therefore, a potential drop $V_{ch}=V_{vari}\sim$ -4 V would be obtained upon submerging the GOG substrate in the saline solution (grounded, $V_{appl}=0$ V). This number is reasonable. In fact, an absolute value of the potential drop $|V_{ch}|$ larger than 7 V causes systematic (i.e. unwanted) nanopore formation.[10] On the other hand, a potential drop $V_{ch}$ at the graphene-electrolyte interface (inside the glass pore) can, in principle, cause various electrochemical processes with electron transfer from graphene to electrolyte. It is most likely that such electron transfer tends to stabilize $V_{ch}$ at -2 V (Fig. 2c), the potential barrier of H$^+$ reduction at graphene cathode (considering the overvoltage). Given the series resistance of the glass pore ($r$=2.5 MΩ) and the charging capacitance $C_{upper}$=60 nF (as $\Delta V_{ch}=\Delta Q_{Gr}/C_{upper}$, $Q_{Gr}$ the charge density in graphene, and $\Delta Q_{Gr}$ the charge transferred during the electrochemical processes), we deduce an effective discharging time constant of $\tau=rC_{upper}$=150 ms. Compared to the fast electrostatic charging of the electric double layer at $C_{lower}$ (100 ns - 1 μs),[15, 16] this discharging process is relatively slow and is therefore presumably due to electron transfer from graphene to electrolyte.

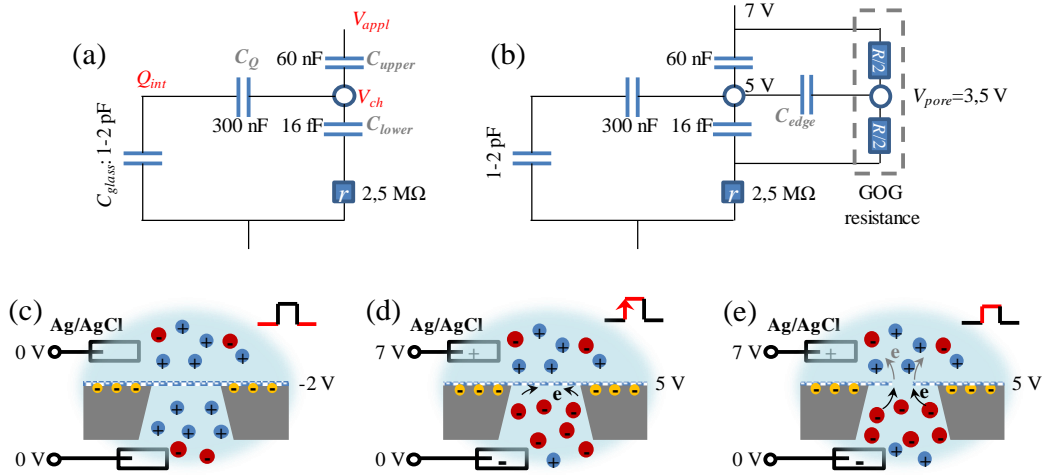

**Figure 2. Mechanisms of GOG nanopore nucleation and formation by electrochemical etching.** (a) Circuit diagram of the GOG system during nucleation. Any applied potential will mainly drop at the graphene/liquid interface inside the glass pore with the smallest capacitance ($C_{lower}$) in the system. (b) Circuit diagram of the GOG system during nanopore formation. The nucleation of the graphene nanopore introduces a possible pathway for the current. As a result, 1) the potential of the electrolyte close to the graphene edge plane will be half of the applied potential: 3.5 V; 2) a potential drop of ~1.5 V=5 V – 3.5 V will be developed across the graphene edges. (c) Scheme of the GOG system when both the "*Cis*" and the "*Trans*" chamber are grounded. (d) and (e) are the schemes of the GOG platform during nanopore nucleation and growth under voltage pulses of 7 V, respectively.

Figure 2d illustrates the charge distribution across the graphene membrane during a voltage pulse. During the first 0.1-1 µs from the start of the 7 V voltage pulse, a voltage drop of $V_{ch}$=5 V (instead of 7 V, due to the offset voltage -2 V in Fig. 2c) develops across $C_{lower}$ (with the smallest capacitance of 16 fF among all the capacitors in the system). The nucleation of the graphene nanopore is taking place under this $V_{ch}$ = 5 V during the following 50 ms time period. Here, the voltage drop at $C_{lower}$ = 5 V has significantly passed the potential barrier of $Cl^-/Cl_2$ and $OH^-/O_2$ oxidization reactions at graphene anode, which occur around 2 V. Consequently, we expect the electrochemical formation of $Cl_2$, $O_2$, and more aggressively, Cl•, O•, and $O_3$• radicals,[17] which can attack carbon atoms in graphene – especially the ones at defect sites in the graphene crystal lattice – and are responsible for atomically sharp nanopore nucleation. Notably, the separation time of 2 s between pulses seems to be sufficient for dissolving any gases formed during the pulse period of 50 ms and stabilizing the system before subsequent pulses, as we did not observe any gas bubbles forming even after up to 200 pulses. 7 V/500 ms pulses, on the other hand, leads to immediate gas bubble formation. As soon as the applied voltage $V_{appl}$ drops from 7 V to 0 V, the actual potential $V_{ch}$ drops to and stabilizes at -2 V again (Fig. 2c). Repeating this process up to 10-20 times normally led to nanopore nucleation with noticeable reduction in the GOG resistance, signaling pore formation, as mentioned above (red triangles, Fig. 1d).

The successful nucleation of a graphene nanopore (7 V, 50 ms, 10 times) introduces a possible pathway of the current. In Figure 2b, given that the resistance of the graphene nanopore ($R>$ 1 GΩ) is dominantly larger than the glass pore resistance ($r$=2-3 MΩ), we expect that half of the applied voltage is distributed at the nanopore: $V_{pore}$ =3.5 V. During the GOG nanopore growth a voltage drop up to 1.5 V occurs across the edges of the graphene pore ($V_{edge} = V_{ch}$ - $V_{pore}$ =5 V - 3.5 V, Fig. 2e), which is sufficient to initiate the electrochemical etching and refreshing effect, as below.

CVD graphene (or even exfoliated graphene) contains defects, including vacancies and grain boundaries. These defective carbon atoms, as well as possible amorphous carbon introduced during chip preparation and storage, are more inclined to be oxidized and attacked in an electrochemical reaction induced by the voltage drops at the graphene/liquid interface. Regrettably, given the extremely small amount of carbon atoms being removed from the graphene membrane during nucleation, it was impossible to directly study the reaction products to resolve the exact mechanisms of nanopore nucleation. However, using liquid atomic force microscopy (AFM) we were able to monitor the in-situ electrochemical etching of graphene (Fig. S4). We were able to trace the electrochemical etching of the graphene along its edges, after applying electrical pulses as in Fig. 3a. Fig. 3b (upper panel) show the AFM image of an as-transferred GOG. This graphene area features exposed edges, represented by the darker yellow areas in Fig. 3b, which can be ascribed to the mechanical damage and/or electrostatic discharge during the assembly of the GOG sample. As shown in the lower panel of Fig. 3b, the electrical pulses attack the graphene edges (white arrows), evident from the widening of the darker regions. That is, defective carbon atoms at the exposed graphene edge are attacked, presumably due to the effective oxidation of carbon into reactive products, which dissolve into the electrolyte solution during the electrochemical etching processes.[18, 19] Naturally, such a solution based *in-situ* electrochemical process produces ultraclean graphene nanopores due to the progressive etching, refreshing and dissolvement compared to TEM sculpt approach.

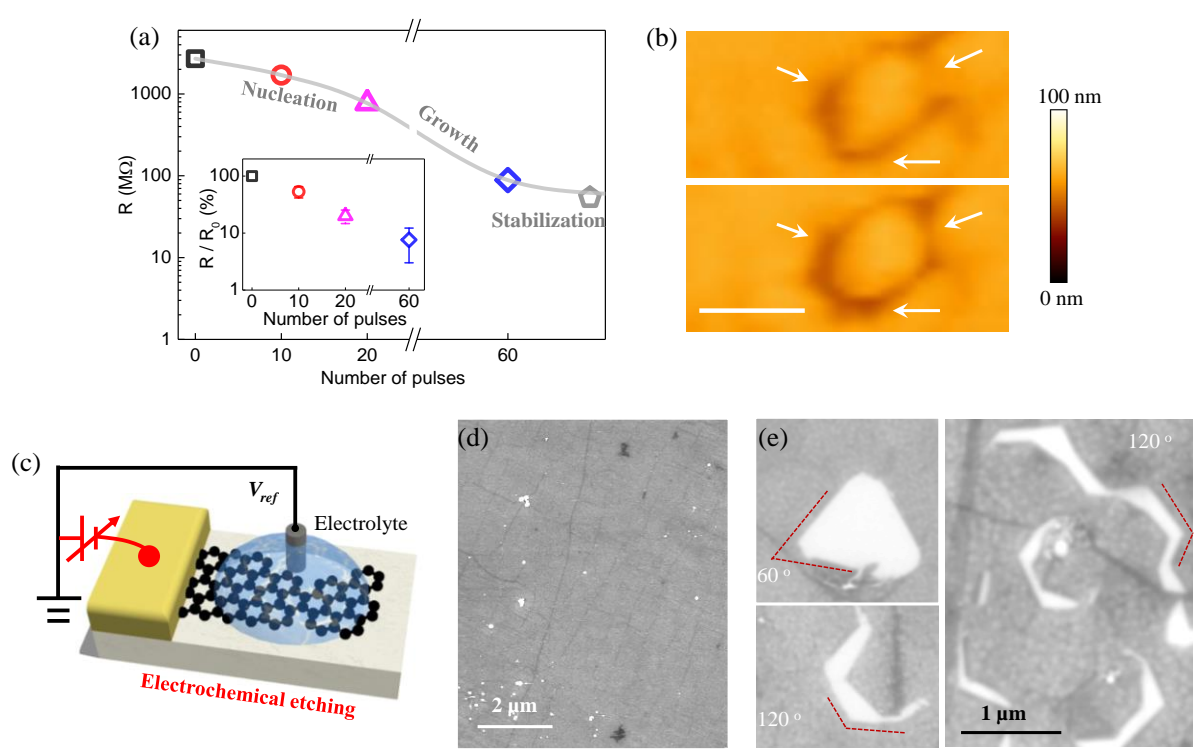

**Figure 3. Ultraclean GOG nanopore formation via electrochemical etching**. (a) The evolution of experimental GOG resistance $R$ vs. number of 7 V/50 ms pulses during electrical fabrication. Modest voltage pulses (5 V/50 ms for 50 times) stabilize the $R$ from 88 MΩ (blue dot) to its final value at 54 MΩ (gray dot). Inset: average relative resistance $R/R_0$ vs. number of pulses for three GOG nanopores, which exhibit similar resistance of ~90 MΩ after experiencing 60 times of 7 V/50 ms pulses. (b) *In-situ* liquid AFM images scanned at the same position during graphene nanostructure enlargement under electric pulses, as indicated by the white arrows (upper panel: before the pulses; lower panel: after the pulses). Scale bar: 50 nm. (c) A scheme of electrochemical etching of CVD graphene transferred on $Al_2O_3$(20 nm)/$SiO_2$(285 nm)/Si substrate and its circuit diagram. (d) An SEM image of as-transferred CVD graphene on the $Al_2O_3$/$SiO_2$/Si substrate. (e) SEM images of the CVD graphene after 30 mins of 1.6 V electrochemical etching in 1 M KCl solution. The exposed hexagonal edges of the CVD graphene (on the insulating $Al_2O_3$/$SiO_2$/Si substrate in white color) suggest that the electrochemical etching process was most likely initiated from its grain boundaries contains active defects.

To further shed light on the mechanism of pore formation, we transferred CVD graphene on an $Al_2O_3$(20 nm)/$SiO_2$(285 nm)/Si substrate (Fig. 3d) and examined electrochemical etching behaviors under similar conditions during nanopore formation. Here, we used a semi-conductive silicon substrate instead of an insulating glass substrate to facilitate SEM imaging. We also applied an additional 20 nm thick atomic layer deposition (ALD) $Al_2O_3$ layer to block any possible leakage current through the substrate.[20] As shown in Figure S5 in the supplementary information, on polycrystalline CVD graphene, the growth started from high-density seeds with hexagonal features (see Fig. S5a). As these hexagonal graphene flakes grew larger laterally and meet each other at the edges, grain boundaries containing vacancy or topological defects formed. In Figure 3c, we applied a DC voltage to the graphene using evaporated gold electrodes or silver paste, to apply a mild voltage of 1.6 V at the graphene/liquid interface via a droplet of 1 M KCl solution (grounded via a reference electrode). This electrochemical condition is similar to the electrical condition during the growth of the graphene nanopore in the previous part of the paper (1.5 V across the edges of the graphene pore, see Fig. 2b and e). After 30 minutes, we were able to etch the CVD graphene as shown in Fig. 3e. The exposed white-colored insulating $Al_2O_3$/$SiO_2$/Si substrate reveals the hexagonal edges of the etched CVD graphene. The fact that the size as well as the hexagonal shape of the etched CVD graphene closely resemble to those of its grain boundaries (see Fig. S5), suggests that the electrochemical etching process was most likely initialized from and progressed along its grain boundaries which contains active defects. It is at these defect sites that a graphene nanopore is preferentially nucleated and grows (50 ms pulses of 5-7 V). Under electrical pulse condition, such electrochemical etching/refreshing process forms atomic sharp pore edge without contaminations, which not only reduce the spatial resolution, but also was supposed as a source for high nanopore noise.[21]

To demonstrate the suitability of these GOG nanopores for single molecule measurements, we have performed DNA translocation experiments (Fig. 4) after the electrical pulse fabrication of the GOG nanopore. We injected λ-DNA on the grounded side of the membrane and applied a negative bias voltage to drive the negatively charged λ-DNA through the GOG nanopore with a 100 mV or 200 mV DC voltage bias via Ag/AgCl reference electrodes (Fig. S6). Threading of the double stranded λ-DNA molecule through the pore causes a transient reduction in the ionic current (conductance) through the pore (Fig. 4a).

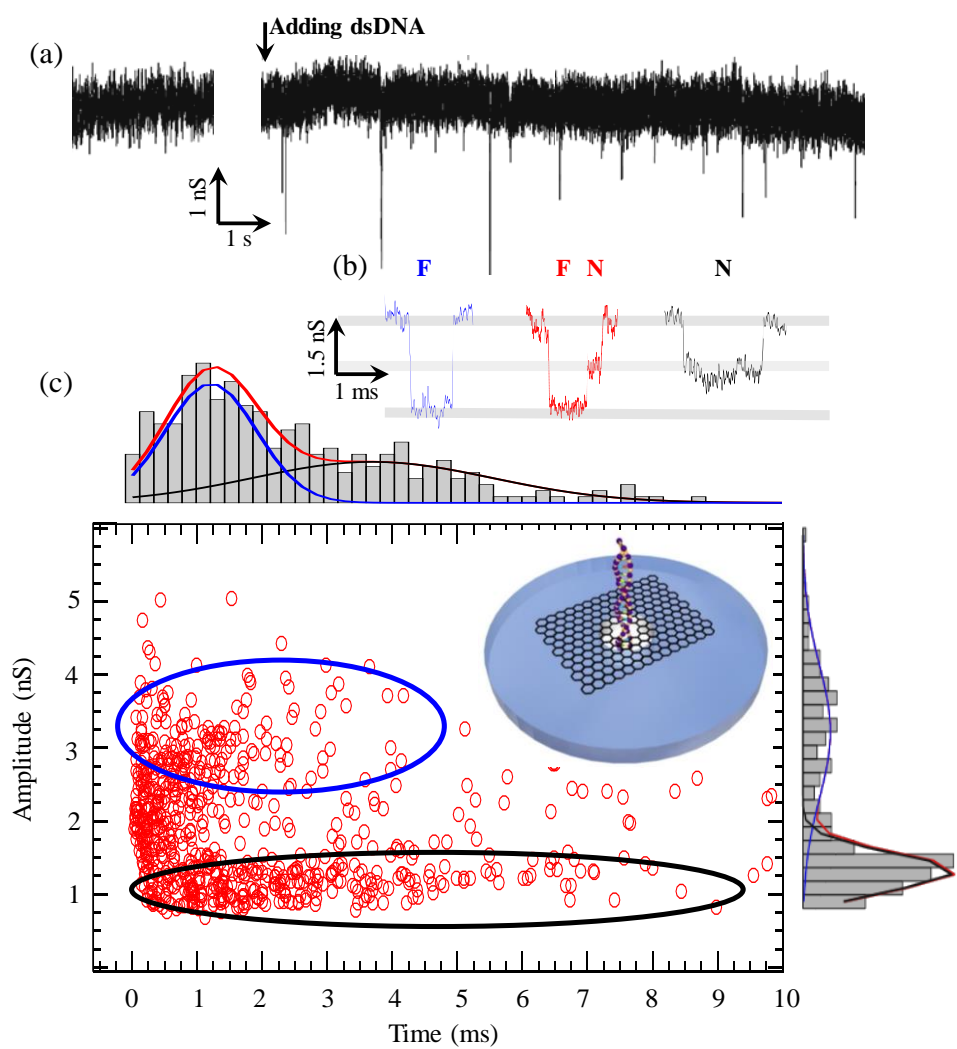

**Figure 4. DNA translocations through a GOG nanopore.** (a) The baseline conductance (left) and blockade events (right) upon addition of 48.5 kb double-stranded λ-DNA across a 16.5 nm (6 MΩ) GOG nanopore. (b) Examples of translocation events of folded (F, blue), partially folded (F+N, red) and nonfolded (N, black) DNA molecules recorded at 200 mV. (c) Scatter diagram of the amplitude of the conductance blockade versus translocation time for λ-DNA translocate through the GOG

nanopores. Insert is a schematic of a DNA molecule translocating through a GOG nanopore. The accompanying histograms for the data are included at the top and the right. Regions of unfolded and folded events are highlighted inside the black and blue circled areas, respectively. Each point in this scatter diagram corresponds to a single translocation event.

These translocation events can be distinguished based on the magnitude of the conductance blockade ($\Delta G$). In Figure 4b, we observed two characteristic reductions in the ionic conductance ($\Delta G$): 1.5 nS and 3 nS, representing the non-folded (N) and folded (F) translocation events, respectively. These $\Delta G$ values for λ-DNA in GOG pores are quite similar in magnitude to that previous measured for graphene pores with diameter of around 20 nm for the same conditions.[2] We measured DNA translocations on 9 graphene nanopores with pore diameters ranging from 10 to 25 nm, and collected good statistics on 4 devices with pore diameters around 20 nm. Based on the conductance measurements, we can now conclude that only a single pore (24 nm for 4 MΩ GOG nanopore) was created during dielectric breakdown, since multiple smaller pores will affect the conductance much more substantially (much larger than 3 nS). The fact that a single GOG nanopore is formed is a clear evidence that the electrochemical etching process is initiated via a fluctuation that randomly selects an active site for the nanopore nucleation and is then limited to that site during the GOG nanopore formation and enlargement. The average translocation time is 2.3±0.9 ms for the 48.5 kb double-stranded λ-DNA (Fig. 4c), corresponding to 0.05 μs/base. This translocation speed is comparable to the previously reported translocation time on λ-DNA in graphene nanopores.[2]

Minimizing noise is of key importance to improve the minimum change in current that can be reliably detected at a given bandwidth. Generally, the noise spectrum in nanopore systems can be divided into a low-frequency regime (f<~1 kHz) and a high-frequency regime (f>~1 kHz). Whereas the low-frequency noise is dominated by pink noise with an 1/*f* dependence;[22]

the high-frequency noise originates from the parasitic capacitance[23] and prevents sampling at high-frequency bandwidths.

Figure 5a provides two typical ionic current traces for GOG nanopores in red and graphene on SiN nanopores in black. Both graphene nanopores were suspended over a pore of 1 micrometer diameter. The traces presented in Figure 5a were both recorded at 200 mV bias voltage and processed following the same protocol (low-pass filtered with an 8-pole Besssel filter at 100 kHz). Strikingly, the GOG nanopore current exhibits much lower noise in the ionic current than the graphene nanopores on the SiN membrane window. In Figure 5b, the corresponding noise spectra reveal one order of magnitude lower high frequency noise for the GOG nanopores than the SiN supported graphene nanopores. At the high-frequency region, the reduction of noise in the GOG nanopores is attributed to the fact that a SiN membrane of tens of nanometer in thickness is replaced by thick glass substrate (thickness of ~100 µm) with low parasitic capacitance, thus significantly reducing the capacitance of the whole support and therefore of the resulting electrical noise consistently.[24]

Similarly, in the low-frequency regime, the amplitude of 1/*f* noise for the GOG nanopore is around one order of magnitude lower than that for graphene nanopores on SiN (Fig. 5b). It is well known that the 1/*f* noise in graphene nanopores fabricated on Si supported SiN membrane is relatively large and is typically two orders of magnitude higher than for silicon nitride nanopores.[25] This large low-frequency 1/*f* noise is most probably due to either possible surface contaminations introduced during preparation/storage or mechanical fluctuations of the free-standing graphene membrane in aqueous solution.[25]

Power spectral density $S_I$ at varying voltage levels for a given graphene pore, shows that $S_I$ curves are bias-voltage dependent due to the large current fluctuations (Fig. 5c). Indeed, the normalized power spectral densities — divide the current power spectral density $S_I$ by the squared current amplitude, $(S_I)/I^2$ — exhibit the same low-frequency noise magnitude and shows no bias-voltage dependent (Fig. 5d). In order to compare the noise levels for different samples under various bias conditions, an equation was deduced $\frac{S_I}{I^2} = \frac{C_{LF}}{f}$, where $C_{LF}$ represents the low-frequency noise amplitude. Here, based on the normalized noise power spectra measurement (Fig. 5d) on a total of 19 GOG chips, the histogram of $C_{LF}$ (Fig. 5e) indicates that the low frequency $1/f$ noise of the GOG nanopore is one order lower than that of electrically fabricated or TEM drilled graphene nanopores on SiN membrane in other studies. That is, the red curves depict log-normal distributions exhibiting an average value of $<C_{LF}>$= $5.8\times10^{-7}$ for GOG nanopores, compared to $<C_{LF}>$= $6.3\times10^{-6}$ for previously reported graphene nanopores in SiN membranes.[25] We ascribe such exceptionally low $1/f$ noise to the ultraclean nature of the electrochemically formed nanopores, as well as to the removal of thermal vibration from the solid support of the glass substrate compared to the free-standing SiN membrane. It was reported that as a nanoscale-microscale electromechanical mixing system, graphene resonator is able to transduce the motion of the coupled free-standing metal electrode with minimal damping.[26, 27] We also investigated the dependence of the $1/f$ noise on the size of the graphene pores. Figure 5f plots the $1/f$ noise as a function of pore diameter. The pore diameters are determined from conductance calculation based on the assumption that the resistance of graphene nanopore is completely dominated by the access resistance $R = \sigma^{-1}\frac{1}{d}$, where $\sigma$ represents the bulk conductivity and $d$ is the pore diameter.[25] Although there is appreciable scatter in the data, there is a correlation between the size and $1/f$ noise where large pores have lower noise which is in accordance with previous studies on SiN membrane supported graphene nanopores.[25] Thus, our noise data clearly demonstrate order-of-magnitude lower current noise in electrochemically formed GOG nanopores, as a consequence of the fabrication method yielding ultraclean nanopores with a favorable reduction in the parasitic capacitive coupling and in the mechanical vibration coupling, making our approach highly promising for potential practical application of graphene nanopores for single molecule detection.

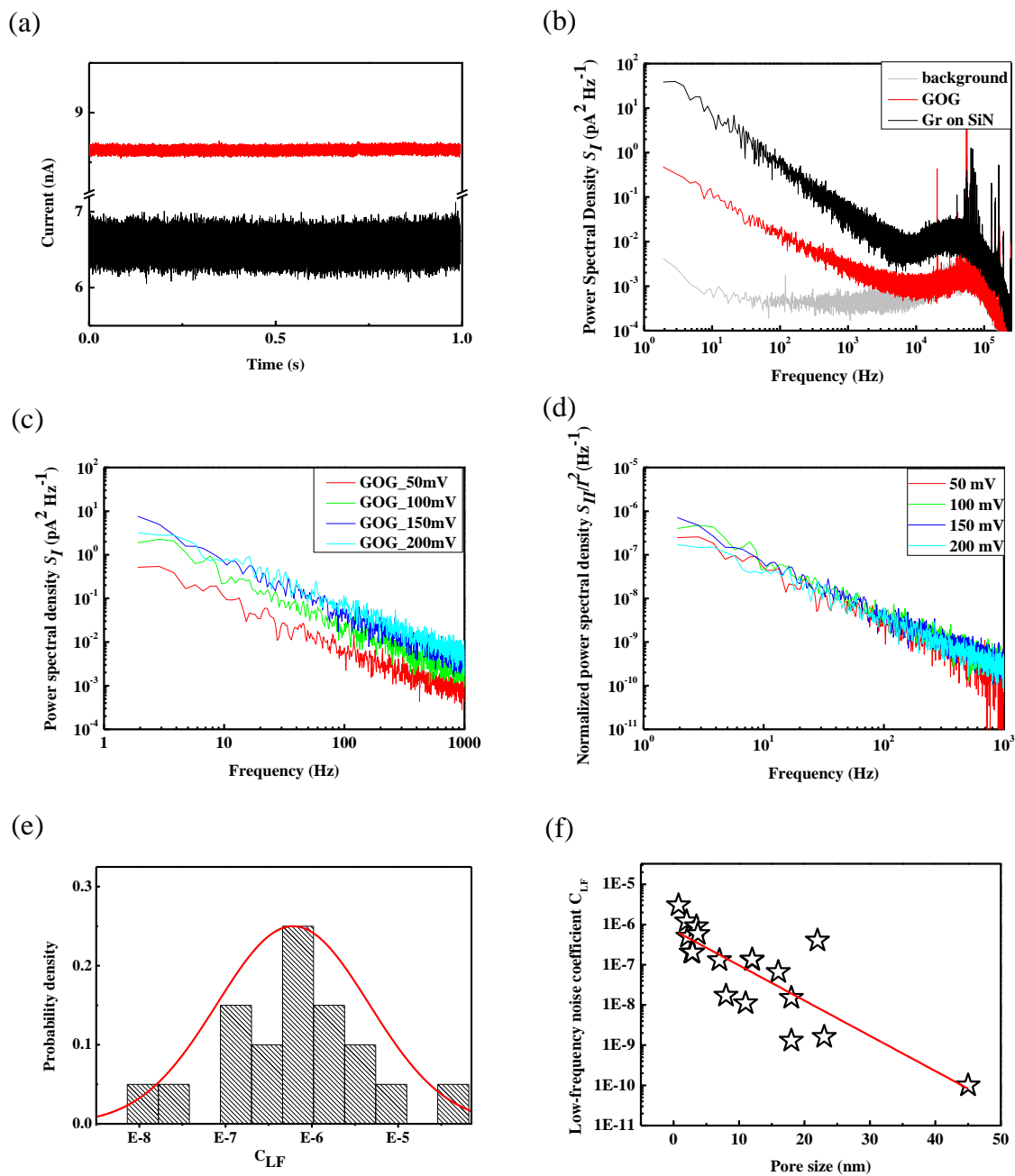

**Figure 5. 1/f noise of GOG nanopore.** (a) Typical current trance of a GOG nanopore (R=23 MΩ, d=4.8 nm in red) and of a graphene nanopore on silicon nitride pore (R=30 MΩ, d=3.8 nm in black) under 200 mV. (b) Spectral densities of the GOG nanopore and of a graphene nanopore on silicon

nitride pore. (c) Spectral densities ($S_I$) at various bias voltages for the GOG nanopore over the 1/$f$ regimes (for graphene pores this band is 1-1000 Hz). (d) The normalized spectral densities ($S_I/I^2$) of the curves in c. A linear fit of these curves yields the low-frequency noise coefficient $C_{LF}$ that represents the magnitude of the 1/$f$ noise. (e) Probability distributions of low-frequency noise coefficients $C_{LF}$ of 19 GOG nanopores. (f) $C_{LF}$ versus pore diameter was plotted double logarithmically. The red line represents a linear fit of the data.

## CONCLUSION

We unraveled the mechanism of ultraclean graphene nanopore/nanogap formation via electrochemical etching based on a graphene-on-glass platform. We conducted *in-situ* liquid AFM studies and demonstrated the enlargement of a graphene nanostructure taking place at its freestanding edges during electrical fabrication. This electrochemical mechanism is strongly supported by our observed electrical etching of CVD graphene on silicon substrate under similar electrochemical conditions. One step further, we confirmed the electrochemical formation of a single graphene nanopore by double-stranded DNA translocation experiments and envision that this electrochemical etching mechanism offers a new strategy to meet the challenges of ultraclean graphene device fabrication at nanometer scale, including nanogaps. Particularly, the ultraclean GOG nanopores with exceptionally low noise level and ease of fabrication under voltage pulse make this a versatile platform to be used for improved nanopore measurement. Furthermore, such platform also featured as a relatively cheap substrate to obtain nanopores in various 2D materials for various nanoelectronics and biomolecule detection applications.

## AUTHOR INFORMATION

**Corresponding Authors**

Wangyang Fu − *School of Materials Science and Engineering, Tsinghua University, No.1 Tsinghua Yuan, Haidian District, 100084 Beijing, China;* E-mail: *fwy2018@mail.tsinghua.edu.cn*

Grégory F. Schneider − *Leiden Institute of Chemistry, Leiden University, Einsteinweg 55, 2333 CC Leiden, Leiden, Netherlands;* E-mail: g.f.schneider@chem.leidenuniv.nl

**Authors**

Xiaoyan Zhang − *Leiden Institute of Chemistry, Leiden University, Einsteinweg 55, 2333 CC Leiden, Leiden, Netherlands*

Pauline M.G. van Deursen − *Leiden Institute of Chemistry, Leiden University, Einsteinweg 55, 2333 CC Leiden, Leiden, Netherlands*

**Notes**

The authors declare that they have no conflict of interest.

**ACKNOWLEDGMENTS**

The authors acknowledge financial support of the European Research Council under the European Union's Seventh Framework Programme (FP/2007-2013)/ERC Grant Agreement no. 335879 project acronym 'Biographene', the Netherlands Organization for Scientific Research (NWO-VIDI 723.013.007, NWO-VENI 722.014.004), the Swiss National Science Foundation (SNSF P300P2_154557, P300P2_164663), European Commission Horizon 2020-Research and Innovation Framework Programme (Marie Sklodowska-Curie actions Individual Fellowship No.749671) and National Natural Science Foundation of China.